\documentclass[twocolumn]{aa}
\usepackage{graphicx}
\usepackage{txfonts}
\def\lsim{\mathrel{\rlap{\lower4pt\hbox{\hskip1pt$\sim$}}
    \raise1pt\hbox{$<$}}}
\def\gsim{\mathrel{\rlap{\lower4pt\hbox{\hskip1pt$\sim$}}
    \raise1pt\hbox{$>$}}}
\begin{document}

   \title{Generalized Chaplygin gas as a unified scenario of dark matter/energy:
		observational constraints
	 }   


   \author{Zong-Hong Zhu
          \inst{1,2,3}
          }

   \offprints{Zong-Hong Zhu}

   \institute{Department of Astronomy, Beijing Normal University,
              Beijing 100875, China
	\and
              National Astronomical Observatories,
              Chinese Academy of Sciences,
              Beijing 100012, China
	\and
	      National Astronomical Observatory,
	      2-21-1, Osawa, Mitaka, Tokyo 181-8588, Japan\\
	      \email{zong-hong.zhu@nao.ac.jp}
             }

   \date{Received 00 00, 0000; accepted 00 00, 0000} 

   \abstract{
   Although various cosmological observations congruously suggest that
     our universe is dominated by two dark components, the cold dark matter
     without pressure and the dark energy with negative pressure, the nature
     and origin of these components is yet unknow.
   The generalized Chaplygin gas (gCg), parametrized by an equation of state,
     $p = -A/\rho_{\rm gCg}^{\alpha}$, was recently proposed to be a candidate
     of the unified dark matter/energy (UDME) scenarios.
   In this work, we investigate some observational constraints on it.
   We mainly focus our attention on the constraints from recent measurements of 
     the X-ray gas mass fractions in clusters of galaxies published by 
	Allen et al. (2002,2003)
     and the dimensionless coordinate distances to type Ia supernovae and
       Fanaroff-Riley type IIb radio galaxies compiled by 
	Daly and Djorgovski (2003).   
   We obtain the confidence region on the two parameters fully characterizing
     gCg, $A_s \equiv A/\rho_{{\rm gCg}0}^{(1+\alpha)}$ and $\alpha$, from
     a combined analysis of these databases, where $\rho_{{\rm gCg}0}$ is the
     energy density of gCg at present.
   It is found that
     $A_s=0.70^{+0.16}_{-0.17}$ and $\alpha=-0.09^{+0.54}_{-0.33}$,
     at a 95\% confidence level, 
    which is consistent within the errors with the standard dark matter + dark
    energy model, i.e., the case of $\alpha = 0$.
   Particularly, the standard Chaplygin gas ($\alpha=1$) is ruled out as a
   feasible UDME by the data at a 99\% confidence level.
   }

   \authorrunning{Zhu, Z.-H.}

   \titlerunning{Generalized Chaplygin gas as a unified scenario of DM \& DE}

   \maketitle

%

\section{Introduction}

Two dark components are invoked to explain the current cosmological
  measurements: the cold dark matter (CDM) without pressure and the
  dark energy (DE) with negative pressure (for a recent review, see
  Peebles \& Ratra 2003).
The first one contributes $\Omega_m \sim 0.3$, and is mainly motivated to
  interprete galactic rotation curves and large scale structure formation
  (e.g. Longair 1998),
while the second one ($\Omega_{\rm DE} \sim 0.7$) provide a mechanism for
  acceleration discovered by distant type Ia supernovae (SNeIa) observations
  (Perlmutter et al. 1998, 1999; Riess et al. 1998, 2001),
  and offset the deficiency of a flat universe, favoured by 
  the measurements of the anisotropy of CMB
        (de Bernardis et al. 2000;
        Balbi et al. 2000,
        Durrer et al. 2003;
        Bennett et al. 2003;
        Melchiorri \& Odman 2003;
        Spergel et al. 2003),
  but with a subcritical matter density parameter $\Omega_m \sim 0.3$,
  obtained from dynamical estimates or X-ray and lensing observations of
  clusters of galaxies(for a recent summary, see Turner 2002).
There are a huge number of candidates for DE in the literature, such as
  a cosmological constant $\Lambda$ 
        (Carroll et al. 1992;
        Krauss and Turner 1995;
	Zhu 1998;
	Sahni 2002;
	Padmanabhan 2003), 
  the so-called ``X-matter"
        (Turner and White 1997;
	Zhu, Fujimoto and Tatsumi 2001;
	Lima and Alcaniz 2002;
	Lima, Cunha and Alcaniz  2003;
	Gong 2004;
	Chen 2004),
  and quintessence
	(Ratra and Peebles 1988;
        Caldwell et al. 1998;
	Sahni and Wang 2000;
	Gong 2002;
	Sahni et al. 2003;
	Padmanabhan and Choudhury 2003)
  etc..    
However, neither CDM nor DE has laboratory evidence for its existence directly.
In this sense, our cosmology depends on two untested entities.
It would be nice if a unified dark matter/energy (UDME) scenario can be found 
  in which
  these two dark components are different manifestations of a single fluid 
	(Padmanabhan \& Choudhury 2002; 
	Wetterich 2002;
	Matos \& Ure\~{n}a-L\'opez 2000).

Recently, the generalized Chaplygin gas (gCg) was proposed as such 
  a unification, which is an exotic fluid with the equation of state 
  as follows
\begin{equation}
p_{\rm gCg} = -A/\rho_{\rm gCg}^{\alpha},
\end{equation}
where $A$ and $\alpha$ are two parameters to be determined.
It was originally suggested by Kamenshchik et al. (2001) with $\alpha = 1$,
  and later on extended by Bento et al. (2002) to gCg.
This simple and elegant model smoothly interpolates between a nonrelativistic
  matter phase ($p=0$) and a negative-pressure dark energy phase 
  ($p=-{\rm const.}$) 
	(Bento et al. 2002;
	Bili\'c et al 2002)
  and moreover it admits a well established brane interpretation
	(Kamenshchik et al. 2001;
	Bento et al. 2002;
	Bili\'c et al 2002).
It is promising to be a candidate of the UDME.
Such a possibility has triggered, quite recently, a wave of interest
  aiming to constrain the gCg model using various cosmological observations,
  such as SNeIa
	(Fabris et al. 2002;
	Makler et al. 2003a,b;
	Bean and Dore 2003;
	Colistete et al. 2003;
	Silva and Bertolami 2003;
	Cunha, Alcaniz and Lima 2004;
	Bertolami et al. 2004),
  the CMB anisotropy measurements
	(Bento et al. 2003a,b;
	Bean and Dore 2003;
	Amendola et al. 2003),
  the gravitational lensing surveys
	(Dev, Jain and Alcaniz 2003, 2004;
	Silva and Bertolami 2003;
	Makler et al. 2003b;
	Chen 2003a,b),
  the X-ray gas mass fraction of clusters
	(Cunha, Alcaniz and Lima 2004;
	Makler et al. 2003b),
  the large scale structure
	(Bili\'c et al. 2002;
	Bean and Dore 2003;
	Multam\"aki et al. 2004),
  and the age measurements of high-$z$ objects
	(Alcaniz, Jain and Dev 2003).	
But the results are disperse and somewhat controversial, with some of them
  claiming good agreement between data and the gCg model while the rest 
  ruling it out as a feasible UDME.

In this work, we shall consider the observational constraints on the parameter
  space of gCg arising from 
  the X-ray gas mass fractions of clusters of galaxies published by 
	Allen et al. (2002, 2003)
  and the dimensionless coordinate distances to SNeIa
  and Fanaroff-Riley type IIb (FRIIb) radio galaxies compiled by 
	Daly and Djorgovski (2003).
We perform a combined analysis of these databases and 
  obtain at a 95\% confidence level,
  $A_s=0.70^{+0.16}_{-0.17}$ and $\alpha=-0.09^{+0.54}_{-0.33}$,
  a parameter range within which the gCg could be a candidate for UDME.
However, the standard Chaplygin gas with $\alpha=1$ is ruled out as a UDME 
  by the data at a 99\% level.
The plan of the paper is as follows.
In the next section, we provide a brief summary of the gCg and basic 
  equations relevant to our work.
Constraints from the X-ray gas mass fractions in galaxy clusters are 
  discussed in section~3.
In section~4 we discuss the bounds imposed by the dimensionless coordinate 
  distances to SNeIa and FRIIb radio galaxies. 
Finally, we present a combined analysis, our conclusion
  and discussion in section~5.


\section{The generalized Chaplygin gas: basic equations}

We consider a flat universe that contains only baryonic matter and the gCg
  (we ignore the radiation components in the universe that are not important
   for the cosmological tests considered in this work). 
Then the Friedmann equation is simply given by 
$H^2 = (8\pi G/3) (\rho_b + \rho_{\rm gCg})$.
Both of the baryonic matter and the gCg components satisfy the relativistic
  energy-momentum conservation equation, 
  $\dot{\rho} + 3{\dot{a}\over a}(p+\rho) =0$, where $a$ is the scale factor
  of the universe and `$\cdot$' stands for the derivative relative to cosmic
  time.
From $p=0$ for the baryonic matter and the equation of state of Eq.(1) for
  the gCg component, we have
\begin{equation}
\rho_{\rm b} = \rho_{{\rm b}0}a^{-3}; \;\
\rho_{{\rm gCg}} = 
 \rho_{{\rm gCg}0} \left(A_s + (1-A_s)a^{-3(1+\alpha)}\right)^{1\over(1+\alpha)}
\end{equation}
where $\rho_{{\rm b}0}$ and $\rho_{{\rm gCg}0}$ are the energy densities of
  the baryonic matter and the gCg at present respectively, and
  $A_s \equiv A/\rho_{{\rm gCg}0}^{1 + \alpha}$ is a substitution of the
  parameter $A$.
The scale factor is related to the observable redshift as $a = 1/(1+z)$.
Now we evaluate the dimensionless coordinate distance, $y(z)$,  
  the angular diameter distance, $D^A (z)$, and
  the luminosity distance, $D^L (z)$, as functions of redshift
  $z$ as well as the parameters of the model.
The three distances are simply related to each other by
  $D^L = (1+z)^2 D^A = (c/H_0) (1+z) y(z)$. 
We define the redshift
  dependence of the Hubble paramter $H$ as $H(z) = H_0 E(z)$,
  where $H_0=100h\,$kms$^{-1}$Mpc$^{-1}$ is the present Hubble constant.
The HST key project result is $h=0.72\pm 0.08$ (Freedman et al. 2001).
Parametrizing the model as $(A_s, \alpha)$, we get $E(z)$ function as
	(Bento et al.2003a,b; Cunha et al. 2004; Alcaniz et al. 2003)
%
\begin{eqnarray}
\label{eq:newE}
E^2(z; A_s, \alpha) & = & \Omega_{\rm b} (1+z)^3 + 
				(1-\Omega_{\rm b}) \cdot \nonumber\\
	& & \left(A_s +(1-A_s)(1+z)^{3(1+\alpha)}\right)^{1\over(1+\alpha)}
\end{eqnarray}
where $\Omega_{\rm b}$ is the density parameter of the baryonic matter 
  component. 
The observed abundances of light elements together with primordial 
  nucleosynthesis give $\Omega_{b}h^{2}=0.0205\pm 0.0018$ 
  (O'Meara et al. 2001).
Then, it is straightforward to show that the distances are given by
%
\begin{eqnarray}
\label{eq:DA}
D^L(z; H_0, A_s, \alpha)& = & (1+z)^2 \cdot D^A(z; H_0, A_s, \alpha)
					\nonumber\\
			& = & {c \over H_0 } (1+z) \cdot y(z; A_s, \alpha)
					\nonumber\\
			& = & {c \over H_0 }(1+z)\cdot\int_{0}^{z} {dz^{\prime} 
				\over E(z^{\prime}; A_s, \alpha) }
\end{eqnarray}
%


\section{Constraints from the X-ray gas mass fraction of galaxy clusters }

As the largest virialized systems in the universe, clusters of galaxies 
  provide a fair sample of the matter content of the whole universe
	(White et. al. 1993).
A comparison of the gas mass fraction of galaxy clusters, 
  $f_{\rm gas} = M_{\rm gas} / M_{\rm tot}$, inferred from X-ray observations, 
  with $\Omega_{\rm b}$ determined by nucleosynthesis can be used to constrain
  the density parameter of the universe $\Omega_m$ directly
	(White \& Frenk 1991;
	Fabian 1991;
	White et. al. 1993;
	White \& Fabian 1995; 
	Evrard 1997; 
	Fukugita, Hogan \& Peebles 1998;
	Ettori \& Fabian 1999).
Sasaki (1996) and Pen (1997) showed that the $f_{\rm gas}$ data of clusters 
  of galaxies at different redshifts can also, in principle, be used to
  constrain other 
  cosmological parameters decribing the geometry of the universe.
This is based on the fact that the measured $f_{\rm gas}$ values for each
  cluster of galaxies depend on
  the assumed angular diameter distances to the sources as 
  $f_{\rm gas} \propto [D^A]^{3/2}$.
The ture, underlying cosmology should be the one which make these measured
  $f_{\rm gas}$ values to be invariant with redshift 
  (Sasaki 1996; Pen 1997; Allen at al. 2003).
However, various uncertainties in previous measurements have seriously
  complicated the application of such methods.

Recently, Allen et al. (2002; 2003) reported precise measurements of the
  $f_{\rm gas}$ profiles for 10 relaxed clusters determined from the
  {\it Chandra} observational data.
Except for Abell 963, the $f_{\rm gas}$ profiles of the other 9 clusters
  appear to have converged or be close to converging with a canonical radius
  $r_{2500}$, which is defined as the radius within which the mean mass 
  density is 2500 times the critical density of the universe at the redshift
  of the cluster (Allen et al. 2002, 2003).
The gas mass fraction values of these 9 clusters are shown in Figure 1.
With the reduced systematic uncertainties, Allen et al. (2002; 2003) 
  successfully applied a method  similar to those proposed by Sasaki (1996) 
  and Pen (1997) to the data and obtained a tight constraint on  
  $\Omega_{\rm m}$ and an interesting constraint on cosmological constant.
We will use this database to constrain the gCg model as a UDME.
Following Allen et al. (2002), we have the model function as
\begin{equation}
f_{\rm gas}^{\rm mod}(z_i;A_s, \alpha) =
      \frac{ b \Omega_b}{\left(1+0.19{h}^{1/2}\right) \Omega_m^{\rm eff}}
  \left[{h\over 0.5}
	\frac{D^A_{\rm{SCDM}}(z_i)}{D^A_{\rm gCg}(z_i;A_s, \alpha)}
		\right]^{3/2}
\end{equation}
where the bias factor $b=0.93\pm 0.05$ (Bialek et al. 2001; Allen et al. 2003)
  is a parameter motivated by gas dynamical simulations, which suggest that
  the baryon fraction in clusters is slightly depressed with respect to the
  Universe as a whole (Bialek et al. 2001). 
The term $(h/0.5)^{3/2}$ represents the change in the Hubble parameter from
  the defaut value of $H_0 = 50 {\rm{km \, s^{-1} \, Mpc^{-1}}}$ and
  the ratio ${D^A_{\rm{SCDM}}(z_i)}/{D^A_{\rm{gCg}}(z_i;A_s, \alpha)}$ 
  accounts for the deviations of the gCg model from the default 
  standard cold dark matter (SCDM) cosmology.
Note that $\Omega_m^{\rm eff}$ is the effective matter density parameter
  (Cunha et al. 2004; Makler et al. 2003b), 
  i.e., the coefficient of the term scaling as
  $(1+z)^3$ in equation (3) when the gCg behaves like dust or equivalently 
  $a \ll 1$.
It is easy to show that, 
$\Omega_m^{\rm eff}=\Omega_{\rm b} + (1-\Omega_{\rm b})(1-A_s)^{1/(1+\alpha)}$.
We should keep in mind that the bias factor value for a gCg model might be
  different from the value given above, which leads to a systematic error
  in this kind of analysis.
Because $b$ linearly scales the X-ray mass fraction, $f_{\rm gas}$, in Eq.(5),
  lowering (raising) it by $\sim 10\%$ would cause the fitting value of
  $\Omega_m^{\rm eff}$ to decrease (increase) by a similar amount.

   \begin{figure}
   \centering
   \includegraphics[width=8.0cm]{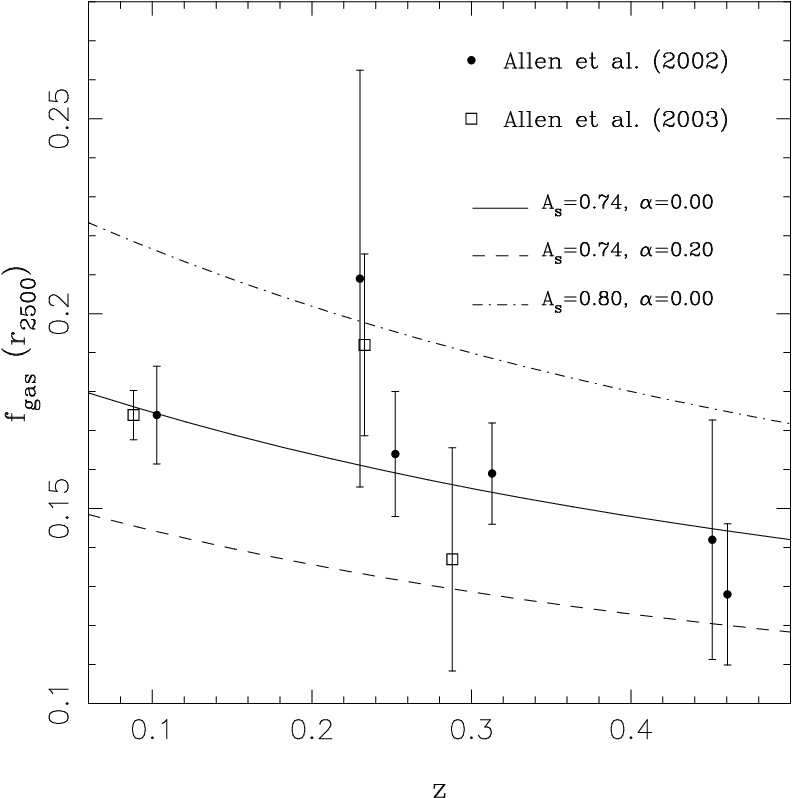}
   \caption{The apparent redshift dependence of the $f_{\rm gas}$ measured 
	    at $r_{2500}$ for 9 clusters of galaxies with convergent 
	    $f_{\rm gas}$ profiles.
	    The error bars are the symmetric root-mean-square $1\sigma$ errors.
	    The solid circles mark the six clusters studied by 
	    Allen et al.(2002), while the empty squares mark the other
	    three clusters published by Allen et al. (2003).
	    The solid curve corresponds our best fit to the gCg model 
	    with $A_s=0.74$, and $\alpha = 0.00$.
           }
   \label{Fig_data1}
    \end{figure}
   \begin{figure}
   \centering
   \includegraphics[width=8.0cm]{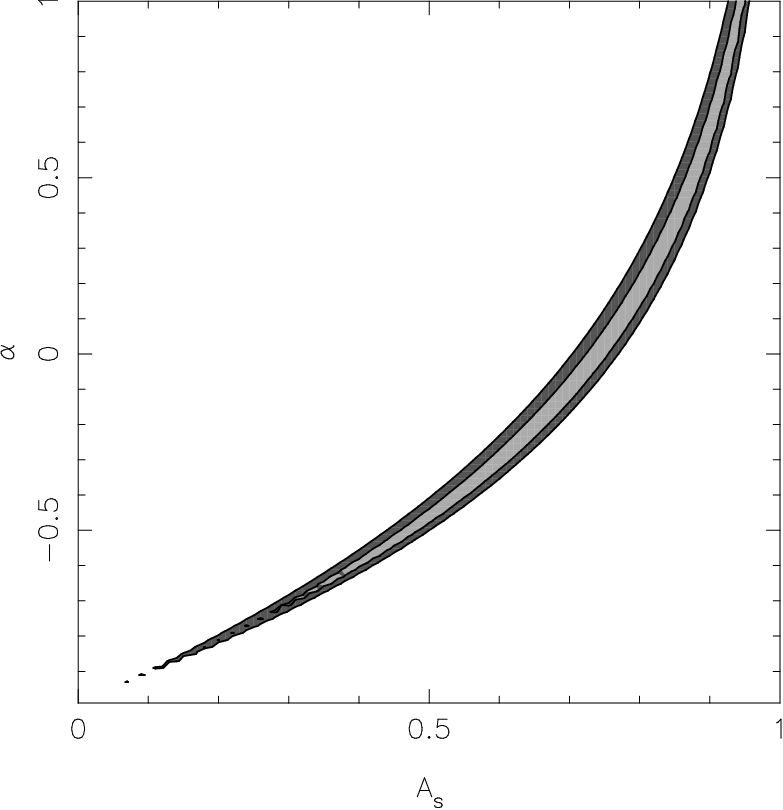}
   \caption{Confidence region plot of the best fit to the $f_{\rm gas}$ of
            9 clusters published by Allen et al. (2002,2003) --
            see the text for a detailed description of the method.
            The 68\% and 95\% confidence levels in the $A_s$--$\alpha$ plane
            are shown in lower shaded and lower $+$ darker shaded areas
            respectively.
           }
   \label{Fig_cont1}
   \end{figure} 

A $\chi^{2}$ minimization method is used to determine the gCg model parameters 
  $A_s$ and $\alpha$ as follows (Allen et al. 2003)
%
\begin{eqnarray}
\label{eq:chi2}
\chi^{2}(A_s, \alpha) & = &    \sum_{i = 1}^{9}
 \frac{\left[f_{\rm gas}^{\rm mod}(z_i;A_s, \alpha) - f_{{\rm gas,o}i}\right]^2}
	{\sigma_{f_{{\rm gas},i}}^2}   \nonumber\\
&&+ \left[\frac{\Omega_bh^{2} - 0.0205}{0.0018}\right]^{2}   +
 \left[\frac{h - 0.72}{0.08}\right]^{2} + \left[\frac{b-0.93}{0.05}\right]^{2},
\end{eqnarray}
where $f_{\rm gas}^{\rm mod}(z_i;A_s, \alpha)$ refers to equation~(5),
  $f_{{\rm gas,o}i}$ is the measured $f_{\rm gas}$ with the defaut SCDM
  cosmology, and $\sigma_{f_{{\rm gas},i}}$ is the symmetric root-mean-square
  errors ($i$ refers to the $i$th data point, with totally 9 data).
The summation is over all of the observational data points.

The results of our analysis for the gCg model are displayed
  in Figure 2.
We show 68\% and 95\% confidence level contours in the ($A_s$,$\alpha$)
  plane using the lower shaded and the lower plus darker shaded areas
  respectively.
The best fit happens at $A_s=0.74$ and $\alpha=0.00$.
Although the data constrain efficiently the parameter plane into a narrow 
  strip, the two parameters, $A_s$ and $\alpha$, are highly degenerate.
This degeneracy can also be seen clearly from the relation,
  $(1-A_s)^{1/(1+\alpha)} = 
	(\Omega_m^{\rm eff}-\Omega_{\rm b})/(1-\Omega_{\rm b})$.
It has been shown that the X-ray gas mass fraction is mostly sensitive to
  $\Omega_m$ no matter what the cosmological model is (Allen et al. 2002, 
  2003; Zhu et al. 2004a,b). 
In our case, a precise determination of $\Omega_m^{\rm eff}$ is expected,
  hence forming a narrow strip in the ($A_s,\alpha$) plane composed of 
  a bundle of curves given by $(1-A_s)^{1/(1+\alpha)} = $const..
In order to determine $A_s$ and $\alpha$ respectively, an independent
  measurement of $A_s$ or $\alpha$ is needed.
We will show that, in the next section, the dimensionless coordinate distances
  to SNeIa and FRIIb radio galaxies are well appropriate for this purpose,
  because the data are only sensitive to $A_s$.


\section{Constraints from the dimensionless coordinate distance data}

Motivated by deriving the expansion rate $E(z)$ and the acceleration rate $q(z)$
  of the universe as functions of redshift,
Daly and Djorgovski (2003) compiled a large database of the 
  dimensionless coordinate distance measurements estimated from the
  observations of SNeIa and FRIIb radio galaxies, 
  and successfully applied it for their purpose.
We will show this sample provides a precise determination of $A_s$, and well
  breaks the degeneracy presented in the X-ray gas mass fraction test.

The database consists in 
  the 54 SNeIa in the ``primary fit C'' used by Perlmutter et al. (1999),
  the 37 SNeIa published by Riess et al. (1998), 
  the so far highest redshift supernova 1997ff presented by Reiss et al. (2001),
  and the 20 FRIIb radio galaxies studied by Daly and Guerra (2002).
The authors used the B-band magnitude-redshift relation,
  $m_B = {\cal M}_B + 5 \log [c(1+z) \cdot y(z)]$, 
  to determine $y(z)$ for each supernova, 
  where ${\cal M}_B \equiv M_B - 5 \log H_0 + 25$ is the 
  ``Hubble-constant-free'' $B$-band absolute magnitude at maximum of a SNIa.
For the 14 supernovae that are present in both the Perlmutter et al. (1999)
  and Riess et al. (1998) samples, we will use their average values
  of $y$ with appropriate error bars (see Table 4 of Daly and Djorgovski 2003).
Therefore we totally have 78 SNeIa data points which are shown as
  solid circles in Figures 3. 
The dimensionless coordinate distances of FRIIb radio galaxies were estimated
  through the method proposed by Daly (1994) 
	(see also Guerra, Daly, and Wan 2000; 
	Podariu et al. 2003; 
	Daly and Djorgovski 2003).
We use their values of $y$ for 20 FRIIb radio galaxies obtained using the best
  fit to both the radio galaxy and supernova data (see Table 1 of
  Daly and Djorgovski 2003), that are shown as empty squares in Figure 3.

   \begin{figure}
   \centering
   \includegraphics[width=8.0cm]{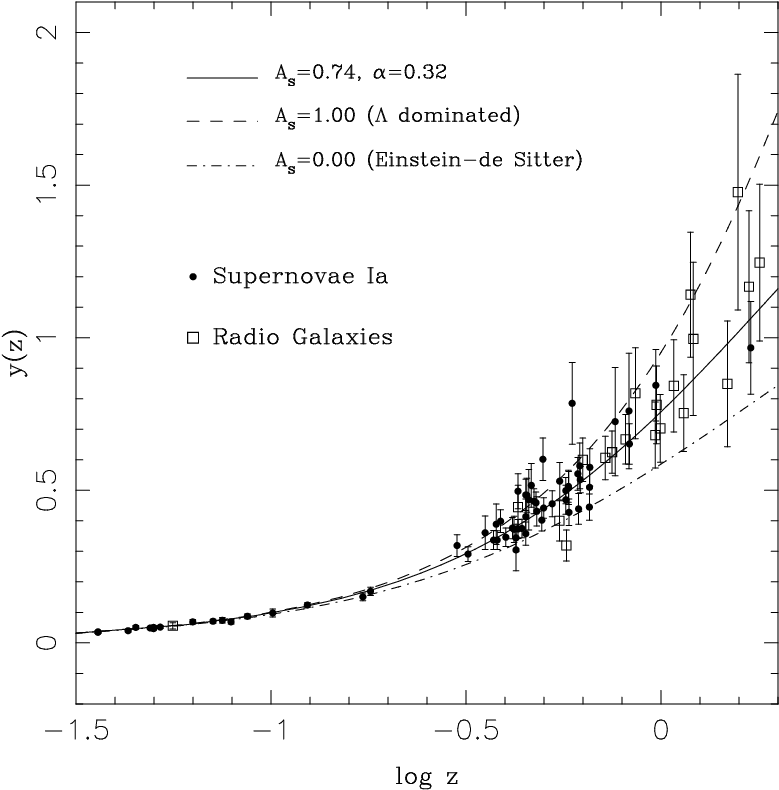}
   \caption{Dimensionless coordinate distances $y(z)$ as a function of $\log z$
	    for 78 type Ia supernovae and 20 FRIIb radio galaxies.
	    The solid circles mark the SNeIa, while the empty squares mark the
	    FRIIb radio galaxies.
	    The solid curve corresponds to our best fit to the total 98 data 
	    points with $A_s=0.74, \alpha=0.32$.
	    The database are taken from Daly and Djorgovski (2003).
	   }
   \label{Fig_data2}		
    \end{figure}
   \begin{figure}
   \centering
   \includegraphics[width=8.0cm]{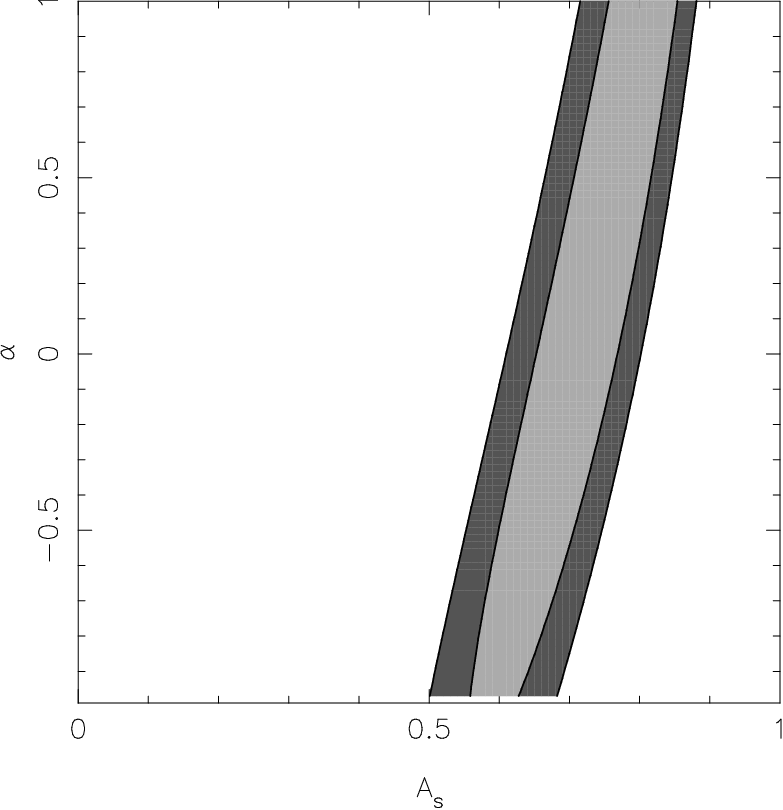}
   \caption{Confidence region plot of the best fit to the
            dimensionless coordinate distances to 78 SNeIa and 20 FRIIb
            radio galaxies compiled by Daly and Djorgovski (2003).
            The 68\% and 95\% confidence levels in the $A_s$--$\alpha$ plane
            are shown in lower shaded and lower $+$ darker shaded areas
            respectively.
           }
   \label{Fig_cont2}
    \end{figure}

We determine the model parameters $A_s$ and $\alpha$ 
  by minimizing
  $\chi^{2}(A_s, \alpha) = \sum_{i=1}^{98}\left[y(z_i;A_s, \alpha) 
	- y_{{\rm o}i}\right]^2/\sigma_i^2$,
where $y(z_i;A_s, \alpha)$ refers to the theoretical prediction from 
  equation~(4), $y_{{\rm o}i}$ is the observed dimensionless coordinate
  distances of SNeIa and FRIIb radio galaxies, and $\sigma_i$ is the 
  uncertainty.

Figure 4 displays the results of our analysis for the gCg model.
We show 68\% and 95\% confidence level contours in the ($A_s$, $\alpha$)
  plane using the lower shaded and the lower plus darker shaded areas
  respectively.
The best fit happens at $A_s=0.74$ and $\alpha=0.32$.
It is clear from the figure, that the dimensionless coordinate distance test
  alone constrains $A_s$ well into a narrow range, but limits $\alpha$ weakly.
However, it is just appropriate for our purpose, to break the degeneracy
  presented in the X-ray gas mass fraction test of last section.
As we shall see in Sec.5, when we combine these two tests,
  we could get very stringent constraints on both
  $A_s$ and $\alpha$, hence test the gCg as a UDME scenario efficiently.


\section{Combined analysis, conclusion and discussion}

Figure 5 displays the results of our combined analysis of the constraints 
  from the X-ray gas mass fractions of galaxy clusters and the dimensionless 
  coordinate distances to SNeIa and FRIIb radio galaxies.
We show 68\%, 95\% and 99\% confidence level contours in the 
  ($A_s$, $\alpha$) plane.  
The best fit happens at $A_s=0.70$ and $\alpha=-0.09$.
As it shown, although
  there is a highly degeneracy between $A_s$ and $\alpha$ in the X-ray mass
  fraction test, and
  the dimensionless coordinate distance test is sensitive to $A_s$ only,
  a combination of the two data sets gives at a 95\% confidence level
  that $A_s=0.70^{+0.16}_{-0.17}$ and $\alpha=-0.09^{+0.54}_{-0.33}$,
  a very stringent constraint on the gCg.
These are the parameter ranges of the gCg permitted 
  by the data as a candidate of UDME, 
  which is consistent within the errors with the standard dark matter + dark
  energy scenario, i.e., the case of $\alpha = 0$.
Particularly, the standard Chaplygin gas with $\alpha=1$ is ruled out as a
   feasible UDME by the data at a 99\% confidence level.  
Using the CMBR power spectrum measurements from BOOMERANG (de Bernardis et al.
  2002) and Archeops (Benoit et al. 2003), together with the SNeIa constraints,
  Bento et al. (2003a) obtained, $0.74 \lsim A_s \lsim 0.85$, and 
  $\alpha\lsim 0.6$, which is comparable with our results.

More recently, Bertolami et al. (2004) analyzed the gCg model in the light of
  the latest SNeIa data (Tonry et al. 2003, Barris et al. 2004).
They considered both the flat and non-flat models.
For the flat case, their best fit values for [$A_s, \alpha$] are given by
  [0.79, 0.999] and [0.936, 3.75] with and without the constraint 
  $\alpha \ge 1$ respectively.
Particularly, up to 68\% confidence level, the $\alpha=0$, i.e., the 
  $\Lambda$CDM case, is clearly excluded, though it is consistent at 95\%
  confidence level (Bertolami et al. 2004).
The authors considered the scenario in which the gCg unified all matter and
  energy components, while in our analysis, only dark matter and dark energy
  are unified as the gCg.
This might be one factor responsible for the difference between their results
  and ours.
Another even more important factor is we make heavy use of the X-ray gas mass
  fraction in clusters, which prefers to $\alpha=0$.
This kind of analysis depends on the assumption that the $f_{\rm gas}$ values
  should be invariant with redshift, which has been criticised by a minority 
  of workers in the field.
For example, a recent comparison of distant clusters observed by XMM-Newton
  and Chandra satellites with available local cluster samples indicate a 
  possible evolution of the $M$--$T$ relation with redshift, i.e., the standard
  paradigm on cluster gas physics need to be revised (Vauclair et al. 2003).
We should keep this point in mind when we compare the results mentioned above.

   \begin{figure}
   \centering
   \includegraphics[width=8.0cm]{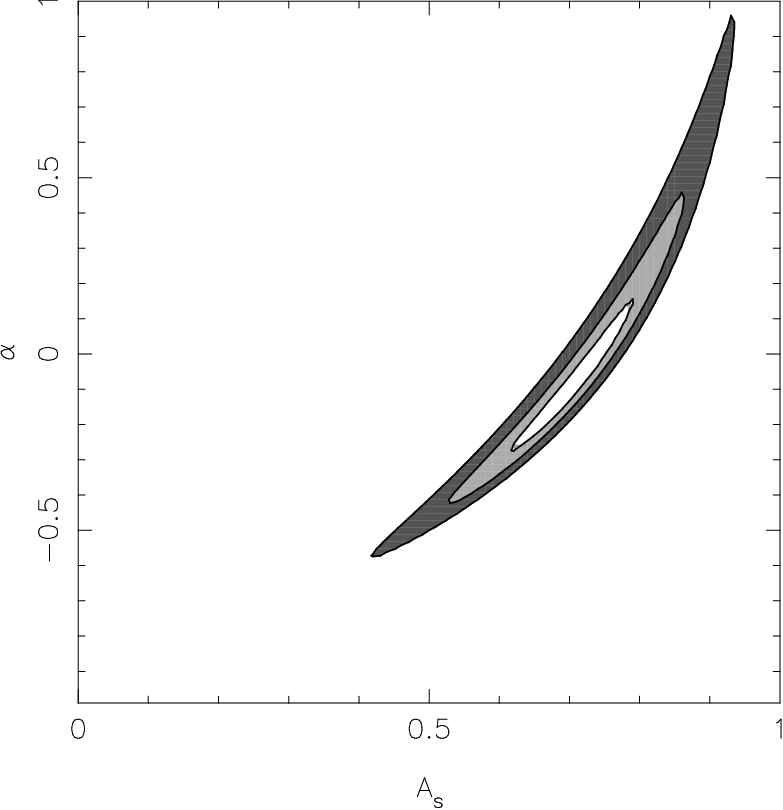}
   \caption{Confidence region plot of the best fit from a combined analysis
	    for the dimensionless coordinate distances to 78 SNeIa and 20 FRIIb
	    radio galaxies (Daly and Djorgovski 2003) and the X-ray gas mass
	    fractions of 9 clusters (Allen et al. 2002, 2003).
	    The 68\%, 95\% and 99\% confidence levels in the 
	    $A_s$--$\alpha$ plane are shown in white, white + lower shaded
	    and white + lower and darker shaded areas respectively. 
           }
   \label{Fig_cont}
    \end{figure}

Besides various dark energy models (see, e.g., Peeble and Ratra 2003), 
  several possible mechanisms without any DE component have been also proposed
  for acceleration of the universe, 
  such as brane world cosmologies 
	(Randall and Sundrum 1999a,b; 
	Alcaniz, Jain and Dev 2002;
	Deffayet, Dvali and Gabadadze 2002;
	Jain, Dev and Alcaniz 2002, 2003),
  and Cardassian expansion model
	(Freese and Lewis 2002;
	Zhu and Fujimoto 2002,2003,2004).
However, it must be more interesting if a UDME can be found in which a single
  fluid plays the role of both CDM and DE.
The generalized Chaplygin gas is such a intriguing candidate, which deserves
  to explore its various observational effects
	(Kamenshchik et al. 2001;
	Bento et al. 2002;2003a,b;
	Bili\'c et al 2002;
	Bean and Dore 2003;
	Cunha, Alcaniz and Lima 2004;
	Makler et al. 2003b).
In this paper we have focused our attention on two observables, 
  the X-ray gas mass fraction and the dimensionless coordinate distance.
We have shown that stringent constraints on the parameters $A_s$ and
  $\alpha$, that completely characterize the scenario,
  can be obtained from the combination analysis of 
  the X-ray mass fractions of galaxy clusters and
  the dimensionless coordinate distances to SNeIa and FRIIb radio galaxies.
It is natually hopeful that, with a more general analysis such as a joint
  investigation on various cosmological observations, one could show clearly
  if this scenario of UDME constitutes a feasible description of our universe.

\begin{acknowledgements}
I would like to thank 
  S. Allen for sending us their compilation of the X-ray mass fraction data and
    his help.
My thanks go to the anonymouse referee for valuable comments and useful
  suggestions, which improved this work very much.
This work was supported by
  the National Natural Science Foundation of China 
	and
  the National Major Basic Research Project of China (G2000077602).
I am also grateful to 
  all TAMA \& LCGT members and the staff of NAOJ for their hospitality 
  and help during his stay.
\end{acknowledgements}

\end{document}